\documentclass{article}
\usepackage{graphicx}
\usepackage{float}
\pdfoutput=1

\usepackage{subcaption}
\usepackage{booktabs} 
\usepackage{multirow}
\usepackage{stfloats}
\usepackage{bm}
\usepackage{amsmath}
\usepackage{amssymb}
\usepackage{xcolor}
\usepackage{makecell}
\usepackage{array}
\usepackage[colorlinks, citecolor=blue]{hyperref}
\newcommand{\PreserveBackslash}[1]{\let\temp=\\#1\let\\=\temp}
\newcolumntype{C}[1]{>{\PreserveBackslash\centering}p{#1}}
\newcolumntype{R}[1]{>{\PreserveBackslash\raggedleft}p{#1}}
\newcolumntype{L}[1]{>{\PreserveBackslash\raggedright}p{#1}}
\usepackage{caption}
\usepackage[capitalize]{cleveref}
\usepackage{cite}
\usepackage{spconf,amsmath,graphicx}

\usepackage{enumitem}
\title{An Empirical Study on the Impact of Positional Encoding in Transformer-based Monaural Speech Enhancement}
\name{\begin{tabular}{@{}c@{}}
Qiquan Zhang$^{1}$, 
Meng Ge$^{2}$,
Hongxu Zhu$^{2,*}$,
Eliathamby Ambikairajah$^{1}$,
Qi Song$^{3}$, 
Zhaoheng Ni$^{4}$, \\
Haizhou Li$^{5,2}$
\end{tabular}
\thanks{\footnotesize{The UNSW is supported by ARC Discovery Grant DP1900102479. The CUHK is supported by 1) NSFC Grant No.~62271432; 2) Guangdong Provincial Key Laboratory of Big Data Computing, CUHK, Shenzhen (Grant No.~B10120210117-KP02); 3) Internal Project Fund from Shenzhen Research Institute of Big Data (Grant No.~T00120220002); 4) Shenzhen Science and Technology Research Fund (Fundamental Research Key Project Grant No.~JCYJ20220818103001002. All models were trained and all experiments were run by the UNSW.} }
\thanks{*Corresponding author}
}

\address{
$^1$School of Electrical Engineering and Telecommunications, University of New South Wales, Australia\\
$^2$Department of Electrical and Computer Engineering, National University of Singapore, Singapore\\
$^3$Alibaba Group, China \, $^4$Meta, United States \\
$^5$The Chinese University of Hong Kong, Shenzhen, China}

\begin{document}
\ninept

\maketitle
%
\begin{abstract}
Transformer architecture has enabled recent progress in speech enhancement. Since Transformers are position-agostic, positional encoding is the de facto standard component used to enable Transformers to distinguish the order of elements in a sequence. However, it remains unclear how positional encoding exactly impacts speech enhancement based on Transformer architectures. In this paper, we perform a comprehensive empirical study evaluating five positional encoding methods, i.e., Sinusoidal and learned absolute position embedding (APE), T5-RPE, KERPLE, as well as the Transformer without positional encoding (No-Pos), across both causal and noncausal configurations. We conduct extensive speech enhancement experiments, involving spectral mapping and masking methods. Our findings establish that positional encoding is not quite helpful for the models in a causal configuration, which indicates that causal attention may implicitly incorporate position information. In a noncausal configuration, the models significantly benefit from the use of positional encoding. In addition, we find that among the four position embeddings, relative position embeddings outperform APEs.






\end{abstract}
\vspace{-0.5em}
\section{Introduction}
\label{sec:intro}


\textcolor{black}{Monaural (single-channel) speech enhancement is the process of separating the clean speech component from a noisy recording to improve the perceived speech quality and intelligibility. It is often deployed as a front-end in a variety of speech processing systems, including speaker verification and identification, automatic speech recognition~\cite{chua2023merlion}, speech coding, and hearing-assistive devices. Conventional unsupervised methods~\cite{loizou2013,logmmse,zhang2019,zhang2019fast} for speech enhancement often exploit the statistical characteristics of speech and noise and derive a clean speech estimator by simplifying the assumptions.~Such methods are incapable of suppressing fast-varying noise sources.}

Over the past decade, supervised speech enhancement based on deep learning has flourished due to increased computational power and the availability of large-scale training data, demonstrating clear superiority over conventional schemes~\cite{wang2018supervised}.~Deep neural networks (DNNs) are exploited to perform the non-linear map from the noisy speech representation to the designed training target.~Neural approaches to speech enhancement could be categorized into waveform domain and time-frequency (T-F) domain schemes. In waveform domain speech enhancement~\cite{kolbaek2020loss}, a popular method is to train a DNN that explicitly adopts an encoder-decoder architecture to directly map the clean speech waveform from the noisy one in an end-to-end manner~\cite{cleanunet}. Neural T-F speech enhancement typically trains a DNN to map the spectral feature of noisy speech to that of clean speech or a spectral mask.~The log-power spectrum (LPS)~\cite{xu2014regression}, magnitude spectrum (MS)~\cite{wang2018supervised}, and complex spectrum (CS)~\cite{tan2019learning} are the most commonly used spectral features. There have been multiple spectral masks developed, including ideal ratio mask (IRM)~\cite{wang2014training}, spectral magnitude mask (SMM)~\cite{wang2014training}, complex IRM (cIRM)~\cite{williamson2015complex}, and phase-sensitive mask (PSM)~\cite{erdogan2015phase}.



To effectively capture the long-range dependencies over time, long short-term memory (LSTM) networks have been explored for speech enhancement~\cite{weninger2015speech,erdogan2015phase,chenlstm}, with substantial improvements over multi-layer perceptrons (MLPs)~\cite{xu2014regression} and better generalization to unseen speakers. Despite the progress, the sequential nature of LSTM networks impairs parallel computation, limiting their use in certain applications. On the other hand, temporal convolution networks (TCNs) utilize stacked 1-D dilated convolution layers to establish a large receptive field, enabling the modeling of long-term correlations~\cite{TCN2018}. Architectures with TCNs have been explored for speech enhancement and showcased impressive performance~\cite{deepmmse,TFA,restcnsa,zhang2021temporal}.

The use of Transformer architectures has driven recent advances in speech enhancement~\cite{sepformerstft,tfaj,ripple,tgsa}. A pivotal component in Transformers is the multi-head self-attention (MHSA) module, which learns the contextual representation by modeling the relevance between each frame with all other frames in parallel. It enables Transformers to learn long-range correlations more efficiently. However, the self-attention module is invariant to the reordering of time frames in an utterance. To help Transformers differentiate time frames at different positions, multiple positional encoding schemes~\cite{pham2020relative} have been explored to inject position information into Transformers. Positional encoding has been utilized in speech enhancement with Transformer architectures~\cite{sepformerstft,tgsa}. However, the exact impact of positional encoding on Transformer speech enhancement remains unclear. Some recent studies~\cite{mhanet,9306484} suggest that positional encoding may not be necessary for speech enhancement, since the noisy spectrogram includes sufficient position information. Hence, a natural question arises: how does positional encoding impact speech enhancement based on Transformers? We perform a systematic empirical study to answer this question across causal and noncausal Transformers, with the most commonly used positional encoding methods. \textcolor{black}{The main contributions of this paper are three-fold:
\vspace{-0.4em}
\begin{itemize} 
    \item To our knowledge, we are the first to conduct a systematic empirical study of how positional encoding affects speech enhancement with Transformers.
    \vspace{-0.3em}
    \item We are the first to study T5-RPE~\cite{T5} and KERPLE~\cite{chi2022kerple} positional encoding methods for speech enhancement.
    \vspace{-0.3em}
    \item Findings highlight the dispensability of positional encoding for the models with a causal configuration and its substantial benefits for models with a noncausal configuration. 
\end{itemize}}


The structure of this paper is organized as follows. In Section~\ref{sec:2}, we introduce neural T-F speech enhancement. Section~\ref{sec:3} elaborates on positional encoding methods. Section~\ref{sec:4} describes speech enhancement with position-aware Transformer. Section~\ref{sec:5} sets up the experiment and analyzes the results. Section~\ref{sec:6} concludes this paper.

\section{Neural Time-Frequency Speech Enhancement}\label{sec:2}

\textcolor{black}{The observed noisy waveform $\bm{x}\!\in\!\mathbb{R}^{1\times T}$ can be formulated as a combination of clean speech $\bm{s}\!\in\!\mathbb{R}^{1\times T}$ and additive noise $\bm{n}\!\in\!\mathbb{R}^{1\times T}$, where $T$ denotes the number of time samples. The time-frequency representation of the noisy mixture is calculated using the short-time Fourier transform (STFT): $X_{l,k}\!=\!S_{l,k}\!+\!N_{l,k}$, where $S_{l,k}$, $N_{l,k}$ and $X_{l,k}$ denote the STFT spectra of clean speech, noise, and noisy mixture, respectively, at the $k$-th frequency bin of the $l$-th time frame. For neural T-F speech enhancement, one typical scheme is to train a DNN model to estimate the clean T-F spectrum $\widehat{S}_{l,k}$ or a T-F mask $\widehat{M}_{l,k}$ from the noisy T-F spectrum at run time. For spectral masking methods, the resulting T-F mask is applied to the noisy spectrum to obtain the clean spectrum: $\widehat{S}_{l,k}\!=\!\widehat{M}_{l,k}\!\cdot\!X_{l,k}$.}

\textcolor{black}{Without the loss of generality, this work adopts the commonly used target magnitude spectrum (MS) $|\widehat{S}_{l,k}|$ and the phase-sensitive mask (PSM)~\cite{erdogan2015phase} to perform speech enhancement to investigate the impact of positional encoding. The PSM is given as:
\begin{equation}\label{PSM}
\setlength{\abovedisplayskip}{2pt}
\setlength{\belowdisplayskip}{2pt}
\text{PSM}[l,k]=\frac{\left|S_{l, k}\right|}{|X_{l, k}|}\cos\left(\phi_{S_{l,k}}-\phi_{X_{l, k}}\right)
\end{equation}
where $|\cdot|$ computes the spectral magnitude, $\phi_{S_{l, k}}$ and $\phi_{X_{l, k}}$ are spectral phases of clean speech and noisy mixture, respectively.}

\vspace{-0.5em}
\section{Positional Encoding}\label{sec:3}

\textcolor{black}{There have been multiple attempts to encode position information into Transformer architectures, which could be grouped into absolute position embedding (APE) and relative position embedding (RPE).}




\vspace{-0.7em}
\subsection{Absolute Position Embedding}
APE represents each absolute position $l$ in a sequence with a unique position vector $\textbf{P}_{l}$ and adds it to the input time frame embedding as the input to the actual Transformer model. The most common choices include fixed and learnable APE. 

\textcolor{black}{\textbf{Sinusoidal Position Embedding}. Vaswani~\textit{et al.}~\cite{transformer} propose to use a sinusoidal function to calculate the position embedding for the original Transformer. In particular, the $j$-th element $\textbf{P}_{l,j}$ of the position embedding for $l$-th position (frame) is given as 
\begin{equation}
\setlength{\abovedisplayskip}{3pt}
\setlength{\belowdisplayskip}{3pt}
\textbf{P}_{l,j}\!=
\begin{cases}
\sin\left(l \cdot 10000^{-{j}/{d_{model}}}\right), & \text{if } j \text{ is even } \\
\cos\left(l \cdot 10000^{-(j-1)/{d_{model}}} \right), & \text{if } j \text{ is odd }
\end{cases}
\end{equation}
where $l\!=\!\left\{1,..,L\right\}$ and $d_{model}$ is the input embedding dimension.}

\textcolor{black}{\textbf{Learnable APE}. Many successful models such as BERT~\cite{bert} and GPT-3 utilize a learnable APE, in which the position embedding $\textbf{P}\!\in\!\mathbb{R}^{L\times d_{model}}$ for each absolute position is learned along with the model.}

\vspace{-0.7em}
\subsection{Relative Position Embedding}

\textcolor{black}{RPE considers relative positions between frames and there have been several RPE methods explored to inject relative position information into Transformers. RPE often works on raw attention matrix with summation before Softmax normalization.}

\textcolor{black}{\textbf{T5-RPE}~\cite{T5} first exploits a log-binning strategy to split the relative positions $i-j$ between frames at positions $i$ and $j$ into a fixed number of buckets, and the same scalar position bias is shared for the positions within the same bucket across Transformer layers. For each attention head, T5-RPE involves a bucket of 32 learnable parameters $\textbf{B}$, and the position bias is assigned as follows:}
\begin{equation}
\setlength{\abovedisplayskip}{2pt}
\setlength{\belowdisplayskip}{2pt}
\textbf{P}_{i,j}\!=\! 
\begin{cases}
\textbf{B}[\min(15,8\!+\!\lfloor\frac{\log ((|i-j|)/8)}{\log(128/8)} \cdot 8\rfloor)\!+\!16], \quad \,\,\, i\!-\!j \! \leq\! -8\\
\textbf{B}[|i-j|+16], \qquad \qquad \qquad \qquad \quad \quad\!{-8}\!<\! i\!-\!j\!<\!0\\ 
\textbf{B}[i-j], \qquad \qquad \qquad \qquad \qquad \qquad \,\,\,\quad 0\!\leq\!i\!-\!j\!<\!8\\ 
\textbf{B}[\min(15,8\!+\!\lfloor\frac{\log ((i-j) / 8)}{\log (128 / 8)} \cdot 8\rfloor)], \ \ \ \, \,\quad \quad \quad \,\,\,\, i\!-\!j\!\geq\!8  
\end{cases}
\end{equation}

\textcolor{black}{\textbf{KERPLE}~\cite{chi2022kerple} employs conditionally positive definite (CPD) kernels to kernelize relative position embeddings, in which the logarithmic variant is better than the power variant. The logarithmic variant is given as:
\begin{equation}\label{kerple}
\setlength{\abovedisplayskip}{2pt}
\setlength{\belowdisplayskip}{2pt}
    \textbf{P}_{i,j} = -r_{1}\cdot \text{log}(1+r_{2}|i-j|)
\end{equation}
where $r_{1}, r_{2}\!>\!0$ are two learnable parameters for each attention head. The position embedding is shared across layers.}

\begin{figure}[!htbp]
\vspace{-0.5em}
\centering
\includegraphics[width=0.91\columnwidth]{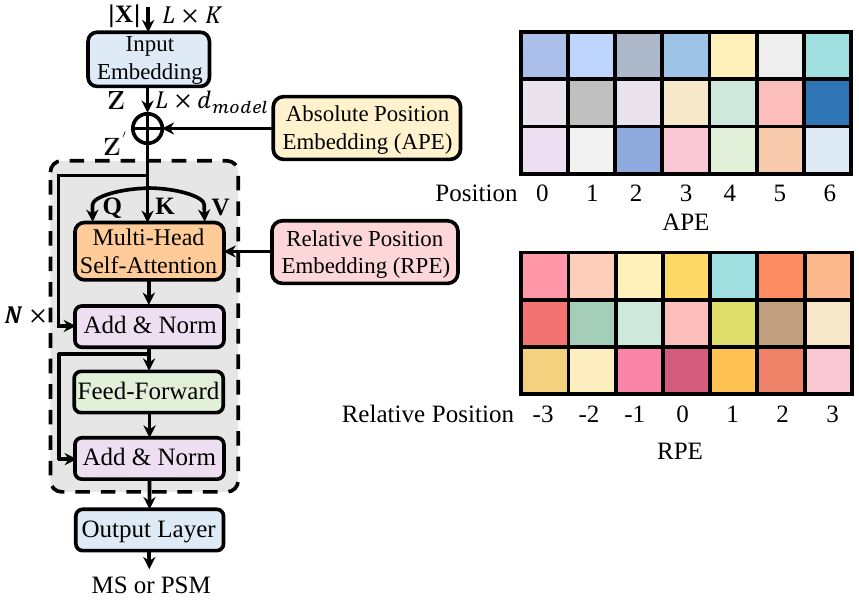}
\vspace{-0.5em}
\caption{\textcolor{black}{Illustration of position-aware Transformer architecture for speech enhancement. $\oplus$ denotes the element-wise summation.}}
\label{fig1}
\vspace{-1.8em}
\end{figure}

\section{Speech Enhancement with Position-aware Transformer}\label{sec:4}

\subsection{Network Architecture}

The overall diagram of the position-aware Transformer for speech enhancement is illustrated in Fig.~\ref{fig1}. It takes as its input the noisy magnitude spectrum (MS) of $L$ frames $|\textbf{X}|\!\in\!\mathbb{R}^{L\times K}$, each frame with $K$ discrete frequency bins. The input is first passed through an embedding layer to obtain the input embedding $\textbf{Z}\!\in\!\mathbb{R}^{L\times d_{model}}$. The embedding layer is a fully-connected (FC) layer where layer normalization (LN) is followed by a ReLU function. APE methods inject the position information by directly adding the position embedding $\textbf{P}$ to $\textbf{Z}$, resulting in $\textbf{Z}^{\prime}$ as the input to $N$-layer Transformer. Each layer consists of an MHSA and a two-layer feed-forward sub-layer. The residual connection followed by LN is used around each sub-layer. The output layer is an FC layer where ReLU and sigmoid activation functions are utilized to predict MS and PSM respectively.

\vspace{-0.9em}
\subsection{Position-Aware Self-Attention}
Unlike APE injecting the position information in the input, RPE incorporates relative position information in self-attention. MHSA involves $H$ attention heads indexed by $h\!\in\!\left\{1,2,..,H\right\}$. For $h$-th head, given a matrix $\textbf{Y}\!\in\!\mathbb{R}^{L\times d_{\text{model}}}$ as the input, it first applies three linear projections to respectively transform the input $\textbf{Y}$ into a set of query, key, and value vectors: $\textbf{Q}_{h}\!=\!\textbf{Y}\textbf{W}^{Q}_{h}$, $\textbf{K}_{h}\!=\!\textbf{Y}\textbf{W}^{K}_{h}$, $\textbf{V}_{h}\!=\!\textbf{Y}\textbf{W}^{V}_{h}$, in which $\{\textbf{W}_{h}^{Q}, \textbf{W}_{h}^{K}\}\!\in \!\mathbb{R}^{d_{\text{model}}\times d_{k}}$, and $\textbf{W}_{h}^{V}\in\!\mathbb{R}^{d_{\text{model}}\times d_{v}}$ are learnable project matrices. $d_{k}$ and $d_{v}$ represents the dimensions of the key and value vectors. Then the scaled dot-product attention is used to calculate the attention matrix and RPE $\textbf{P}_{h}\!\in\!\mathbb{R}^{L\times L}$ is injected via the element-wise summation before Softmax normalization:
\begin{equation}
\setlength{\abovedisplayskip}{5pt}
\setlength{\belowdisplayskip}{5pt}
    \textbf{A}_{h} = \text{Softmax}\left(\textbf{M}\odot\left[{\textbf{Q}_{h}\textbf{K}_{h}^\top\!/\!}{\sqrt{d_{k}}} + \textbf{P}_{h} \right]\right)\textbf{V}_{h}
\end{equation}
where $d_{k}\!=\!d_{v}\!=\!d_{model}\!/\!{H}$ and $\textbf{M}\!\in\!\{1,0\}^{L\times L}$ is an attention mask that controls what context a frame can attend to. The attention scores corresponding to $\textbf{M}_{i,j}\!=\!0$ are masked out and that corresponding to $\textbf{M}_{i,j}\!=\!1$ are involved. The operation $\odot$ is defined as:
\begin{equation}
\setlength{\abovedisplayskip}{2pt}
\setlength{\belowdisplayskip}{2pt}
 \left(\textbf{M} \odot \textbf{E} \right)_{i j}= \begin{cases}
  -\infty & \text{ if } \, \textbf{M}_{i j}=0 \\
  \textbf{E}_{i j} & \text{ if } \, \textbf{M}_{i j}=1 
 \end{cases}
\end{equation}
The outputs of all $H$ attention heads are then concatenated and transformed with a linear projection:
\begin{equation}
\setlength{\abovedisplayskip}{5pt}
\setlength{\belowdisplayskip}{5pt}
    \text{MHSA}\left(\textbf{Q}, \textbf{K}, \textbf{V}\right) = \text{Concat}\{\textbf{A}_{1},...,\textbf{A}_{H}\}\textbf{W}^{O}
\end{equation}
where $\textbf{W}^{O}\!\in\!\mathbb{R}^{d_{\text{model}}\times d_{\text{model}}}$. The output of MHSA is further processed by a two-layer feed-forward network (FFN)~\cite{transformer}.

\begin{figure}[!hbtp]
\vspace{-1.0em}
\centering
\begin{subfigure}[t]{0.37\columnwidth}
\centerline{\includegraphics[width=0.95\columnwidth]{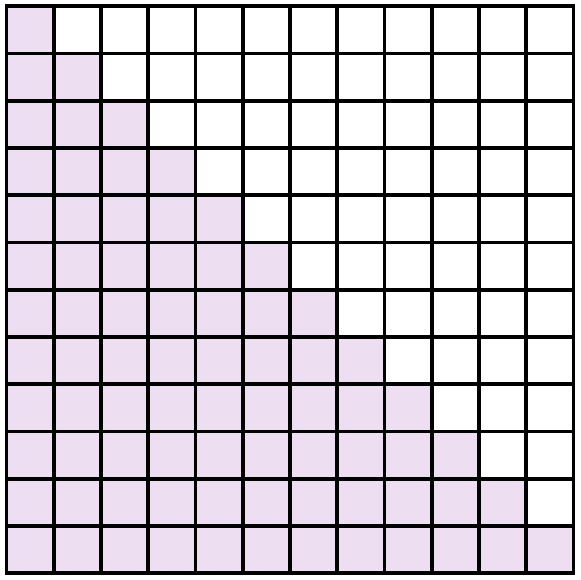}}
\caption{causal self-attention}
\label{fig2:1}
\end{subfigure}
\hspace{3mm}
\begin{subfigure}[t]{0.37\columnwidth}
\centerline{\includegraphics[width=0.95\columnwidth]{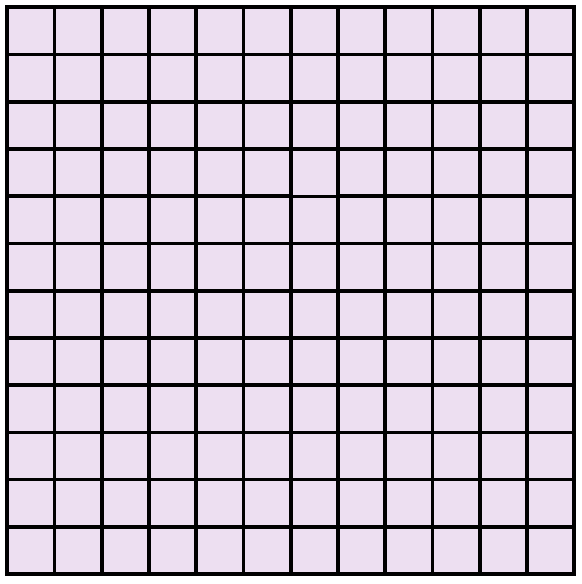}}
\caption{noncausal self-attention}
\label{fig2:2}
\end{subfigure}
\vspace{-0.7em}
\caption{Illustration of (a) the causal self-attention and (b) the noncausal (full) self-attention (with sequence length $L=12$).}
\vspace{-1.0em}
\label{fig2}
\end{figure}

\textcolor{black}{\textbf{Causal and Noncausal self-attention}. Here, we consider both causal and noncausal configurations. Fig.~\ref{fig2} illustrates an example of causal and noncausal self-attention patterns. The blank squares ($\textbf{M}_{i,j}\!=\!0$) denote that the corresponding attention scores are masked out. The colored squares ($\textbf{M}_{i,j}\!=\!1$) denote that the attention scores are involved in MHSA. In the causal models, a causal attention mask (Fig.~\ref{fig2}\,(a)) is exploited to mask out the upper triangular scores of the attention matrix to block the interactions from future frames, ensuring causality. In noncausal self-attention (Fig.~\ref{fig2}\,(b)), the attention scores from all the frames are involved. }

\section{EXPERIMENTS}\label{sec:5}
\subsection{Data and Feature Extraction}

\textcolor{black}{The clean speech clips for training are collected from LibriSpeech \textit{train-100} corpus~\cite{panayotov2015librispeech}, comprising $28,539$ clips in total, with approximately 100 hours of speech. The noise clips are drawn from multiple datasets~\cite{tfaj}, i.e., the noise data of the MUSAN datasets~\cite{snyder2015musan}, the RSG-10 dataset~\cite{steeneken1988description}, the Environmental Noise dataset~\cite{saki2016automatic,saki2016smartphone}, the colored noise set~\cite{deepmmse}, the UrbanSound dataset~\cite{Urban}, the QUT-NOISE dataset~\cite{dean2010qut}, and the Nonspeech dataset~\cite{hu2010tandem}. We exclude four noise sources from the noise data for evaluation: the \textit{street music} from the UrbanSound dataset, and \textit{voice babble}, \textit{factory welding}, and \textit{F16} from the RSG-10 dataset. The resulting noise set contains $6,809$ noise clips. For validation experiments, $1,000$ noise and clean speech clips are randomly picked and mixed to build a validation set of $1,000$ noisy clips, where each clean speech clip is degraded by a random section from one noise clip at a random SNR level (sampled from -10 to 20 dB, in 1 dB steps). For each of the four noises for testing, we randomly pick ten clean speech clips from LibriSpeech \textit{test-clean-100} corpus and degrade each clip by a random section cut from the noise source at SNRs of \{-5, 0, 5, 10, 15\} dB, resulting in 200 noisy clips for evaluation~\cite{tfaj}. All signals are sampled at 16 kHz. We use the STFT spectral magnitude as the input to networks, which is computed using a 512 (32 ms) long square-root Hann window with a hop length of 256 (16 ms).}

\vspace{-1.3em}
\subsection{Implementations}
\textcolor{black}{To study the role of positional encoding in speech enhancement, we employ a Transformer architecture without positional encoding (No-Pos) as a base and consider causal and noncausal systems. The Transformer architecture includes $N\!=\!4$ stacked Transformer layers, with the parameter configurations as: $d_{\text{model}}\!=\!256$, $H\!=\!8$, and $d_{f\!f}\!=\!1024$. Our experiments consider four commonly used positional encoding methods, i.e., Sinusoidal-APE~\cite{transformer}, Learned-APE~\cite{bert}, T5-RPE~\cite{T5}, and KERPLE~\cite{chi2022kerple}.}

\textcolor{black}{We use the mean square error (MSE) as the objective function and adopt mask approximation to learn PSM. We employ Adam optimizer~\cite{Adam} for gradient optimization and
set its hyperparameters as in~\cite{ripple}, i.e., $\epsilon\!=\!1\text{e}{-9}$, $\!\beta_{1}\!=\!0.9$, and $\!\beta_{2}\!=\!0.98$. We utilize gradient clipping to cut gradient values to the range of $[-1,\, 1]$. The models are trained for 150 epochs where the mini-batch size is 10 utterances. The noisy mixtures are produced dynamically at training time. In particular, we mix each clean speech clip picked for one batch with a random segment of one random noise clip at an SNR level randomly sampled from $\left\{-10\leq s \leq 20\,| s\in \mathbb{Z}\right\}$ (dB). The warm-up strategy~\cite{transformer,tfaj} is used to adjust the learning rate: $lr\!=\!d_{\text{model}}^{-0.5}\cdot \textrm{min} \left(\text{n\_itr}^{-0.5}, \text{n\_itr} \cdot \text{wup\_itr}^{-1.5}\right)$, where $\text{n\_itr}$ and $\text{wup\_itr}$ are the number of iterations and warm-up stage iterations, respectively. We follow~\cite{tfaj} and set $\text{wup\_itr}$ to $40\,000$.}

\begin{figure}[!h]
\vspace{-0.8em}
\centering
\begin{subfigure}[!htbp]{0.48\columnwidth}\centerline{\includegraphics[width=\columnwidth]{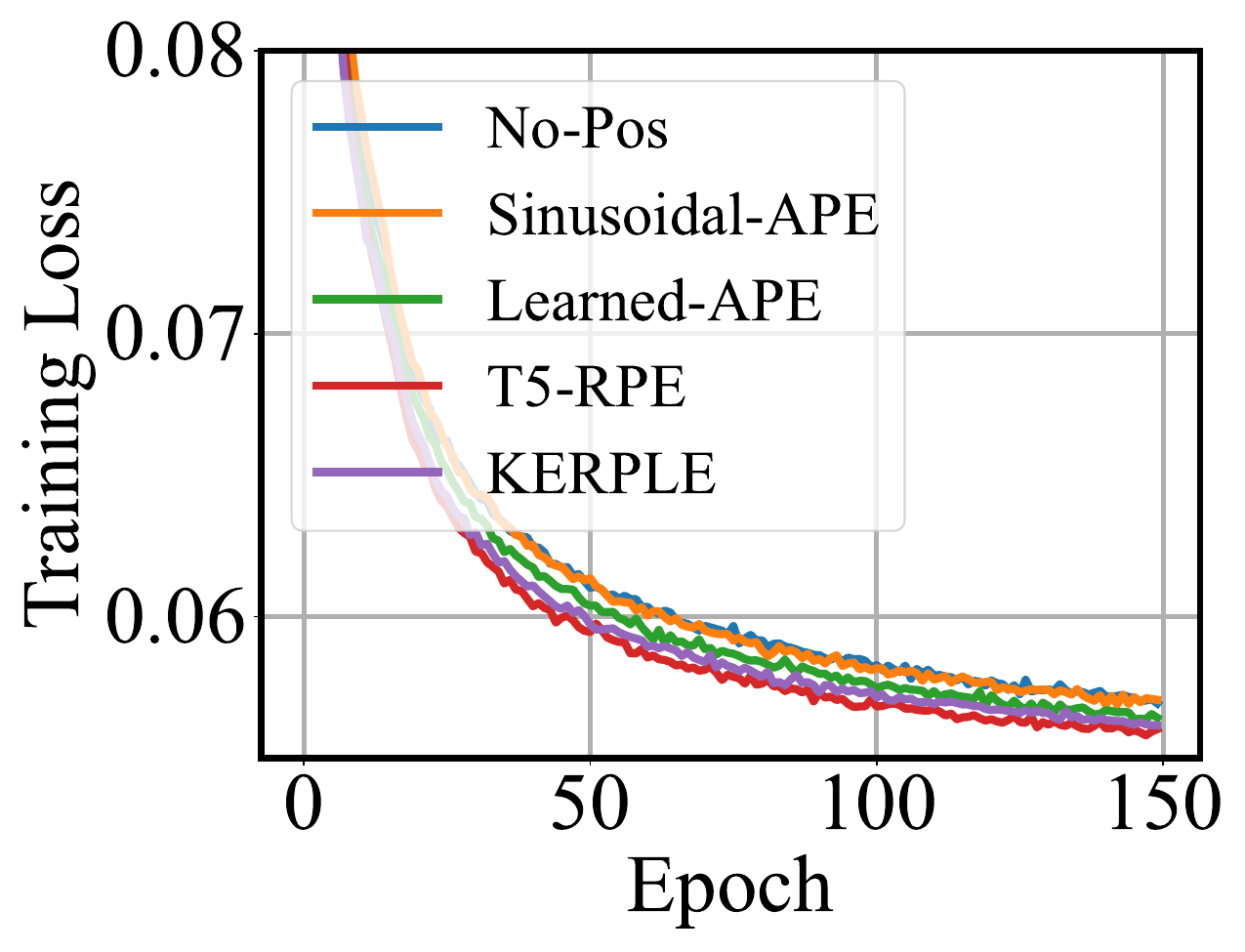}}
\label{fig3:1}
\end{subfigure}
\begin{subfigure}[!htbp]{0.48\columnwidth}
\centerline{\includegraphics[width=\columnwidth]{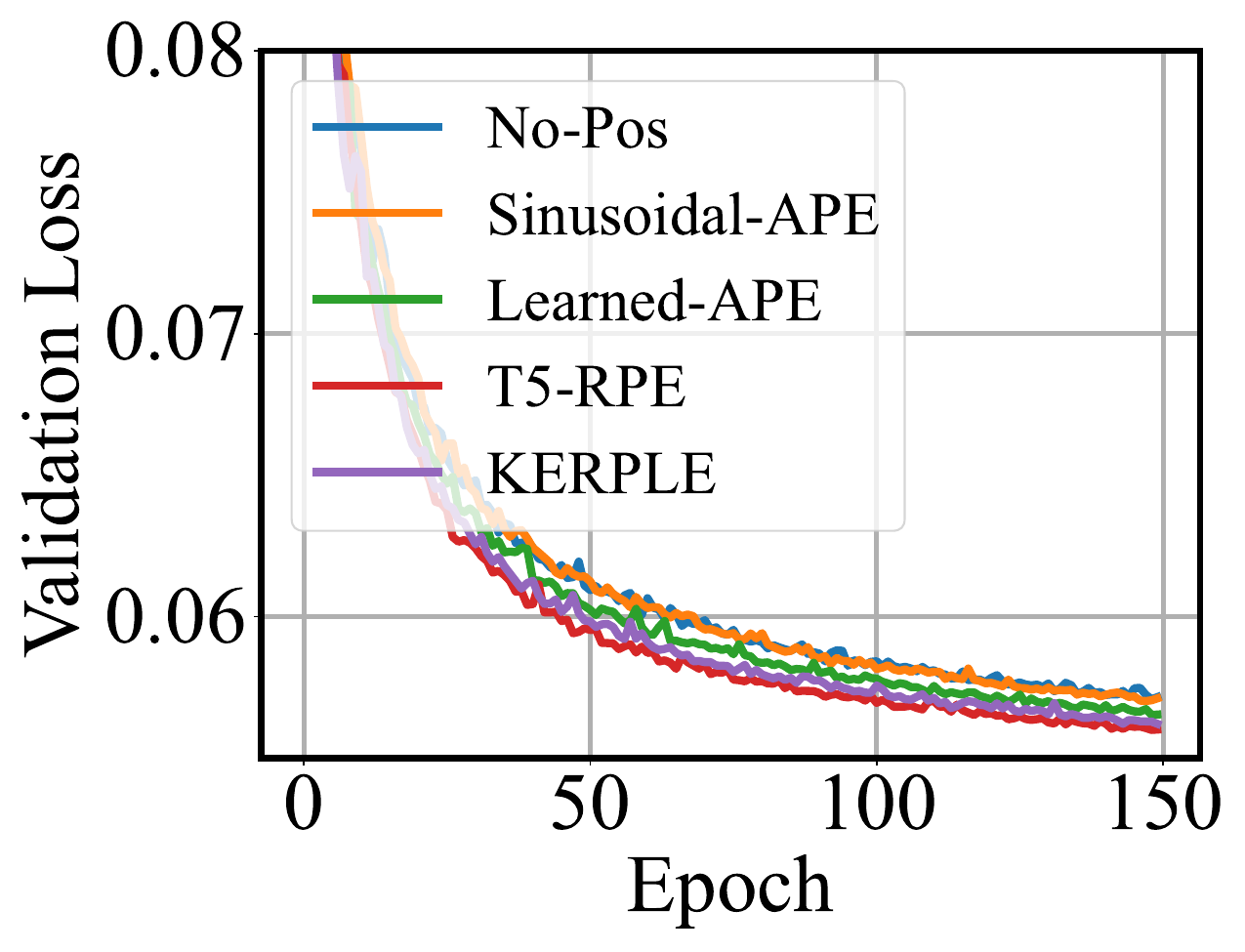}}
\label{fig3:2}
\end{subfigure}
\vspace{-0.8em}
\caption{The training and validation loss in causal configuration.}
\label{fig3}
\vspace{-1.0em}
\end{figure}

\begin{figure}[!h]
\vspace{-1em}
\centering
\begin{subfigure}[!htbp]{0.48\columnwidth}\centerline{\includegraphics[width=\columnwidth]{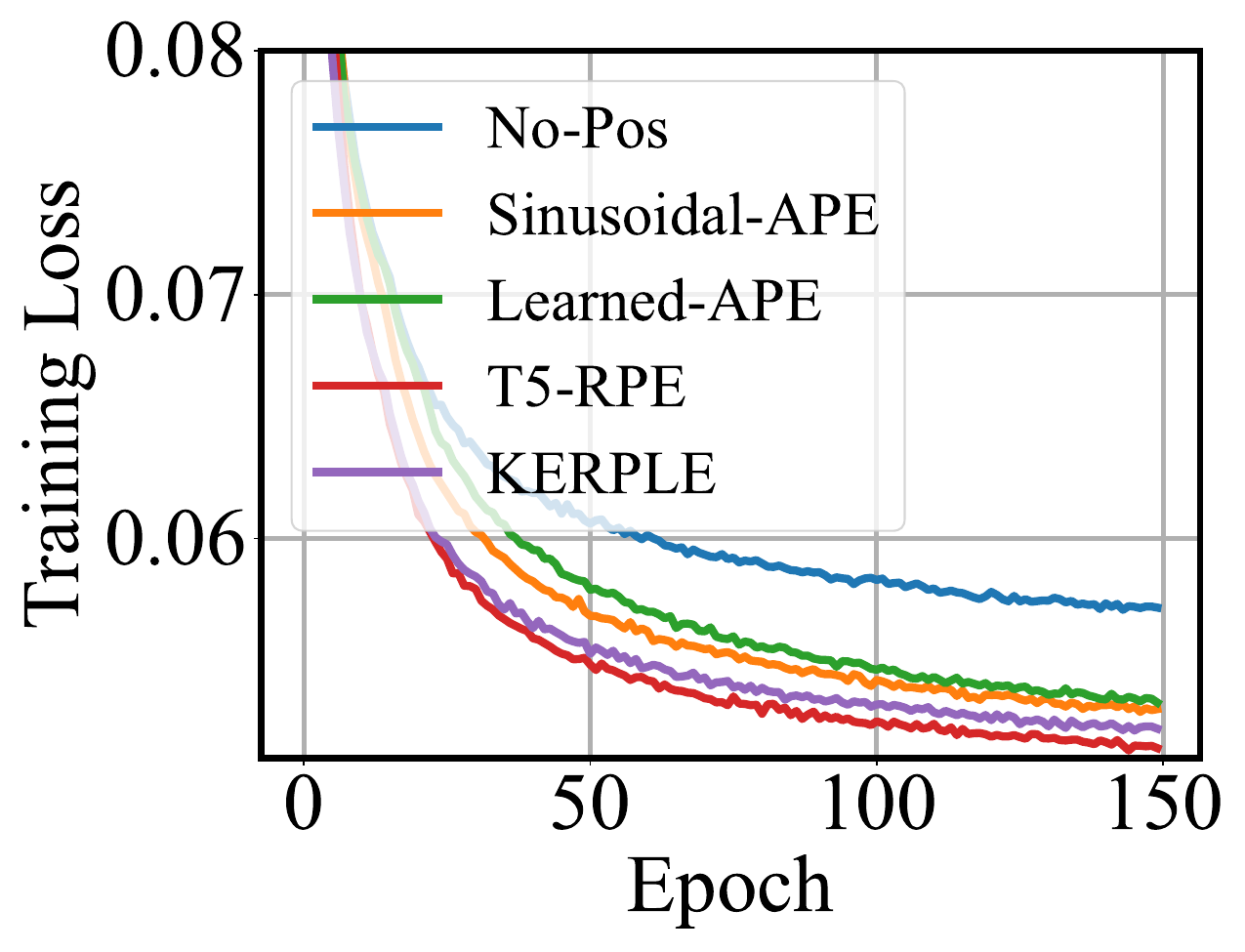}}
\label{fig4:1}
\end{subfigure}
\begin{subfigure}[!htbp]{0.48\columnwidth}
\centerline{\includegraphics[width=\columnwidth]{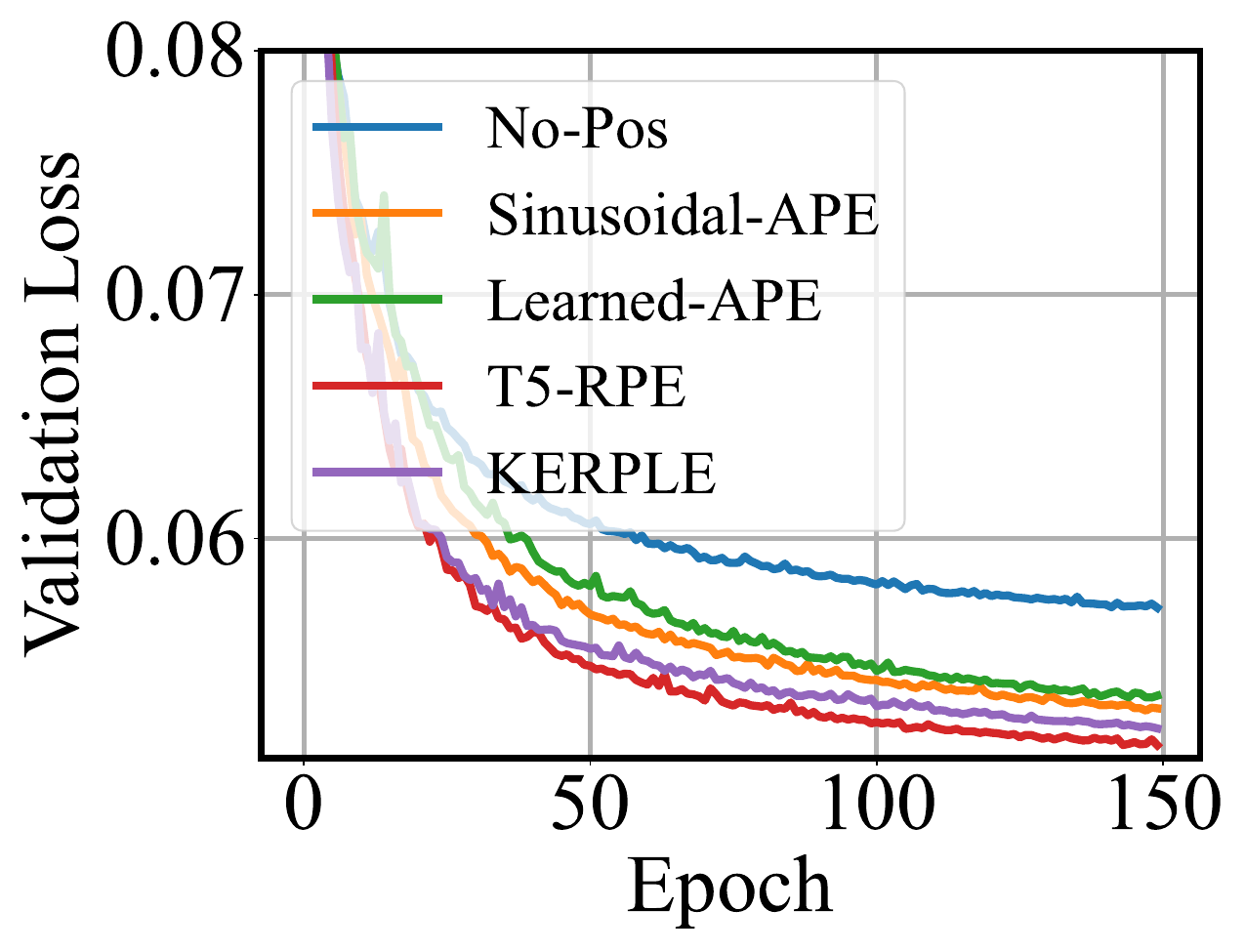}}
\label{fig4:2}
\end{subfigure}
\vspace{-1em}
\caption{The training and validation loss in noncausal configuration.}
\label{fig4}
\vspace{-1.9em}
\end{figure}

\vspace{-0.5em}
\subsection{Training \& Validation Loss}

Fig.~\ref{fig3}-\ref{fig4} illustrate the training and validation loss curves of different position embedding methods in causal and noncausal configurations, respectively, with the PSM as the training target. We can easily observe that the impact of the position embedding is quite different for causal and noncausal systems. It can be found that the benefits of position embeddings to the causal system are quite limited. We explain this by analyzing the function of the causal mask. Given a frame at position $m$, the causal mask only allows the model to attend to the previous $m$ frames to update the representation at position $m$, which is not permutation invariant and may implicitly inject position information. In a noncausal configuration, given a frame at any position, all the frames are used to update the representation, which is unaware of the position. In contrast, the use of position embeddings consistently achieves substantially lower training and validation losses than the noncausal No-Pos model. Among these four position embedding methods, overall, RPE methods (T5-RPE and KERPLE) are superior to APE methods (Sinusoidal- and learned-APE).

\begin{table*}[!t]
    \centering
    \footnotesize
    \def\arraystretch{0.93}
    \setlength{\tabcolsep}{5.5pt}
    \setlength{\abovetopsep}{0pt}
    \setlength\belowbottomsep{0pt} 
    \setlength\aboverulesep{0pt} 
    \setlength\belowrulesep{0pt}
    \caption{PESQ scores of different position embedding methods. The best PESQ scores under each SNR level are in boldface.}
    \label{tabwer}
    \vspace*{0.03in}
    \scalebox{0.92}{
    \begin{tabular}{l|c|c|ccccc|ccccc}
        \toprule
        \multirow{3}{*}{\makecell[c]{\textbf{Methods}}} & 
        \multirow{3}{*}{\makecell[c]{\textbf{Types}}} &
        \multirow{3}{*}{\makecell[c]{\textbf{Causality}}} & 
        & & \textbf{MS} & & & &  & \textbf{PSM} &  & \\
        \cmidrule{4-13}
        &  &  & \multicolumn{5}{c|}{Input SNR (dB)} 
            & \multicolumn{5}{c}{Input SNR (dB)} \\
        \cmidrule{4-13}
        &  &  & -5 & 0 & 5 & 10 & 15 
              & -5 & 0 & 5 & 10 & 15 \\
        \midrule
        Noisy & -- & -- 
        & 1.28 & 1.50 & 1.84 & 2.21 & 2.58
        & -- & -- & -- & -- & -- \\
        \midrule
        No-Pos &  -- & \multirow{5}{*}{\textbf{Yes}} & 
        \textcolor{black}{1.78} & \textcolor{black}{2.25} & \textcolor{black}{2.62} & \textcolor{black}{2.92} & \textcolor{black}{3.18} &
        
        \textcolor{black}{1.79} & \textcolor{black}{2.27} & \textcolor{black}{2.68} & \textcolor{black}{3.04} & \textcolor{black}{3.35} \\
        
        Sinusoidal-APE~\cite{transformer} &  \multirow{2}{*}{APE} &  & 
        \textcolor{black}{1.78} & \textcolor{black}{2.26} & \textcolor{black}{2.63} & \textcolor{black}{2.94} & \textcolor{black}{3.19} & 
        
        \textcolor{black}{1.78} & \textcolor{black}{2.26} & \textcolor{black}{2.68} & \textcolor{black}{3.04} & \textcolor{black}{3.34} \\
        
        Learned-APE~\cite{bert} &  &  
        & \textcolor{black}{1.78} & \textcolor{black}{2.25} & \textcolor{black}{2.62} & \textcolor{black}{2.93} & \textcolor{black}{3.19}& 
        
        \textcolor{black}{1.79} & \textcolor{black}{2.28} & \textcolor{black}{2.70} & \textcolor{black}{3.04} & \textcolor{black}{3.36} \\
        
        T5-RPE~\cite{T5} & \multirow{2}{*}{RPE} &  & 
        \textcolor{black}{\textbf{1.80}} & \textcolor{black}{\textbf{2.29}} & \textcolor{black}{\textbf{2.66}} & \textcolor{black}{\textbf{2.96}} & \textcolor{black}{\textbf{3.21}} &
        
        \textbf{1.82} & \textbf{2.30} & 2.71 & \textbf{3.06} & \textbf{3.37} \\
        KERPLE~\cite{chi2022kerple}
        & & & 
        \textcolor{black}{\textbf{1.80}} & \textcolor{black}{2.26} & \textcolor{black}{2.65} & \textcolor{black}{\textbf{2.96}} & \textcolor{black}{\textbf{3.21}} & 
        
        \textcolor{black}{1.81} & \textcolor{black}{2.30} & \textcolor{black}{\textbf{2.72}} & \textcolor{black}{3.06} & \textcolor{black}{3.36} \\

        \midrule
        \midrule
        No-Pos &  -- & \multirow{5}{*}{\textbf{No}} & 
        \textcolor{black}{1.76} & \textcolor{black}{2.22} & \textcolor{black}{2.60} & \textcolor{black}{2.91} & \textcolor{black}{3.18} &
        
        \textcolor{black}{1.78} & \textcolor{black}{2.26} & \textcolor{black}{2.66} & \textcolor{black}{3.01} & \textcolor{black}{3.32} \\
        
        Sinusoidal-APE~\cite{transformer} &  \multirow{2}{*}{APE} &  & 
        
        \textcolor{black}{2.00} & \textcolor{black}{2.43} & \textcolor{black}{2.79} & \textcolor{black}{3.09} & \textcolor{black}{3.33} &
        
        \textcolor{black}{1.95} & \textcolor{black}{2.43} & \textcolor{black}{2.82} & \textcolor{black}{3.16} & \textcolor{black}{3.46} \\
        
        Learned-APE~\cite{bert} &  &  
        & \textcolor{black}{1.99} & \textcolor{black}{2.43} & \textcolor{black}{2.77} & \textcolor{black}{3.07} & \textcolor{black}{3.32}& 
        
        \textcolor{black}{1.92} & \textcolor{black}{2.41} & \textcolor{black}{2.82} & \textcolor{black}{3.16} & \textcolor{black}{3.45} \\
        
        T5-RPE~\cite{T5} & \multirow{2}{*}{RPE} &  & 
        \textcolor{black}{\textbf{2.05}} & \textcolor{black}{\textbf{2.51}} & \textcolor{black}{\textbf{2.86}} & \textcolor{black}{\textbf{3.14}} & \textcolor{black}{\textbf{3.38}} &
        
        \textcolor{black}{\textbf{1.99}} & \textcolor{black}{\textbf{2.49}} & \textcolor{black}{\textbf{2.90}} & \textcolor{black}{\textbf{3.24}} & \textcolor{black}{\textbf{3.53}} \\
        KERPLE~\cite{chi2022kerple}
        & & & 
        \textcolor{black}{\textbf{2.05}} & \textcolor{black}{2.49} & \textcolor{black}{2.83} & \textcolor{black}{3.11} & \textcolor{black}{3.35} & 
        
        \textcolor{black}{1.97} & \textcolor{black}{2.45} & \textcolor{black}{2.85} & \textcolor{black}{3.19} & \textcolor{black}{3.48} \\
               
        \toprule
    \end{tabular}}
    \label{pesq}
    \vspace{-1.0em}
\end{table*}

\begin{table*}[!t]
    \vspace{-0.5em}
    \centering
    \footnotesize
    \def\arraystretch{0.95}
    \setlength{\tabcolsep}{4.8pt}
    \setlength{\abovetopsep}{0pt}
    \setlength\belowbottomsep{0pt} 
    \setlength\aboverulesep{0pt} 
    \setlength\belowrulesep{0pt}
    \caption{ESTOI (in \%) scores of different position embedding methods. The best ESTOI scores under each SNR level are in boldface.}
    \label{tabwer}
    \vspace*{0.03in}
    \scalebox{0.92}{
    \begin{tabular}{l|c|c|ccccc|ccccc}
        \toprule
        \multirow{3}{*}{\makecell[c]{\textbf{Methods}}} & 
        \multirow{3}{*}{\makecell[c]{\textbf{Types}}} &
        \multirow{3}{*}{\makecell[c]{\textbf{Causality}}} & 
        & & \textbf{MS} & & & &  & \textbf{PSM} &  & \\
        \cmidrule{4-13}
        &  &  & \multicolumn{5}{c|}{Input SNR (dB)} 
            & \multicolumn{5}{c}{Input SNR (dB)} \\
        \cmidrule{4-13}
        &  &  & -5 & 0 & 5 & 10 & 15 
              & -5 & 0 & 5 & 10 & 15 \\
        \midrule
        Noisy & -- & -- &
        27.91 & 42.14 & 57.21 & 71.11 & 82.22
        & -- & -- & -- & -- & -- \\
        \midrule
        No-Pos &  -- & \multirow{5}{*}{\textbf{Yes}} & 
        \textcolor{black}{44.62} & \textcolor{black}{61.41} & \textcolor{black}{74.04} & \textcolor{black}{82.29} & \textcolor{black}{87.50} &
        
        \textcolor{black}{43.47} & \textcolor{black}{61.19} & \textcolor{black}{74.95} & \textcolor{black}{84.27} & \textcolor{black}{90.17} \\
        
        Sinusoidal-APE~\cite{transformer} &  \multirow{2}{*}{APE} &  & 
        \textcolor{black}{44.90} & \textcolor{black}{61.98} & \textcolor{black}{74.40} & \textcolor{black}{82.58} & \textcolor{black}{87.51} & 
        
        \textcolor{black}{43.68} & \textcolor{black}{61.40} & \textcolor{black}{75.19} & \textcolor{black}{84.39} & \textcolor{black}{90.24} \\
        
        Learned-APE~\cite{bert} &  &  
        & \textcolor{black}{44.73} & \textcolor{black}{61.95} & \textcolor{black}{74.51} & \textcolor{black}{82.57} & \textcolor{black}{87.73}& 
        
        \textcolor{black}{43.63} & \textcolor{black}{61.54} & \textcolor{black}{75.34} & \textcolor{black}{84.41} & \textcolor{black}{90.24} \\
        
        T5-RPE~\cite{T5} & \multirow{2}{*}{RPE} &  & 
        \textcolor{black}{\textbf{45.84}} & \textcolor{black}{\textbf{62.56}} & \textcolor{black}{\textbf{74.81}} & \textcolor{black}{82.75} & \textcolor{black}{\textbf{87.79}} &
        
        \textcolor{black}{\textbf{44.42}} & \textcolor{black}{\textbf{62.21}} & \textcolor{black}{\textbf{75.73}} & \textcolor{black}{\textbf{84.75}} & \textcolor{black}{\textbf{90.42}} \\
        KERPLE~\cite{chi2022kerple}
        & & & 
        \textcolor{black}{45.50} & \textcolor{black}{62.39} & \textcolor{black}{\textbf{74.81}} & \textcolor{black}{\textbf{82.76}} & \textcolor{black}{87.68} & 
        
        \textcolor{black}{44.17} & \textcolor{black}{61.98} & \textcolor{black}{75.67} & \textcolor{black}{84.69} & \textcolor{black}{90.37} \\

        \midrule
        \midrule

        No-Pos &  -- & \multirow{5}{*}{\textbf{No}} &
        
        \textcolor{black}{42.24} & \textcolor{black}{59.99} & \textcolor{black}{73.46} & \textcolor{black}{82.51} & \textcolor{black}{87.85} &
        
        \textcolor{black}{42.35} & \textcolor{black}{60.51} & \textcolor{black}{74.61} & \textcolor{black}{84.21} & \textcolor{black}{90.17} \\
        
        Sinusoidal-APE~\cite{transformer} &  \multirow{2}{*}{APE} &  &
        
        \textcolor{black}{49.14} & \textcolor{black}{65.38} & \textcolor{black}{76.86} & \textcolor{black}{84.28} & \textcolor{black}{88.76} &
        
        \textcolor{black}{47.64} & \textcolor{black}{65.09} & \textcolor{black}{77.62} & \textcolor{black}{85.85} & \textcolor{black}{91.18} \\
        
        Learned-APE~\cite{bert} &  &  
        & \textcolor{black}{48.42} & \textcolor{black}{64.91} & \textcolor{black}{76.80} & \textcolor{black}{84.28} & \textcolor{black}{88.91}& 
        
        \textcolor{black}{46.81} & \textcolor{black}{64.21} & \textcolor{black}{77.25} & \textcolor{black}{85.61} & \textcolor{black}{91.00} \\
        
        T5-RPE~\cite{T5} & \multirow{2}{*}{RPE} &  & 
        \textcolor{black}{\textbf{51.92}} & \textcolor{black}{\textbf{67.72}} & \textcolor{black}{\textbf{78.43}} & \textcolor{black}{\textbf{85.24}} & \textcolor{black}{\textbf{89.45}} &
        
        \textcolor{black}{\textbf{49.84}} & \textcolor{black}{\textbf{66.98}} & \textcolor{black}{\textbf{78.92}} & \textcolor{black}{\textbf{86.63}} & \textcolor{black}{\textbf{91.58}} \\
        KERPLE~\cite{chi2022kerple}
        & & & 
        \textcolor{black}{50.32} & \textcolor{black}{66.66} & \textcolor{black}{77.95} & \textcolor{black}{84.97} & \textcolor{black}{89.33} & 
        
        \textcolor{black}{48.00} & \textcolor{black}{65.48} & \textcolor{black}{78.00} & \textcolor{black}{86.21} & \textcolor{black}{91.34} \\
        \toprule
    \end{tabular}}
    \label{estoi}
    \vspace{-1.0em}
\end{table*}

\vspace{-0.5em}
\subsection{Results and Discussion}


\textcolor{black}{We use PESQ~\cite{recommendation2001perceptual}, ESTOI~\cite{jensen2016algorithm}, and three composite metrics (CSIG, CBAK, and COVL)~\cite{hu2007evaluation} to assess enhanced speech signals.}




\textcolor{black}{Tables~\ref{pesq}-\ref{estoi} showcase the PESQ and ESTOI scores of different position embedding methods across the five SNR levels, respectively. We first clearly observe that position embedding shows quite different impacts on Transformer speech enhancement in causal and noncausal configurations. Specifically, it can be found that in the causal configuration, position embeddings provide quite limited or no performance gains over No-Pos. For instance, causal No-Pos with PSM archives the same or better PESQ scores (Table~\ref{pesq}) than Sinusoidal-APE under five SNR levels. In the case of -5 dB SNR (on MS), compared to causal No-Pos, Sinusoidal- and Learned-APE show the same PESQ scores, and KERPLE and T5-RPE only provide 0.02 gains. However, we can find that position embeddings substantially improve the PESQ and ESTOI scores of noncausal No-Pos, under all five SNR conditions. In the 5 dB SNR case (on MS), for instance, Sinusoidal- and Learned-APE, T5-RPE, and KERPLE provide 0.19 and 3.40\%, 0.17 and 3.34\%, 0.26 and 4.97\%, and 0.23 and 4.49\% PESQ and ESTOI gains over noncausal No-Pos, respectively. In addition, the results show the superiority of RPE methods over APE methods. Table~\ref{composite} lists the average CSIG, CBAK, and COVL scores of different position embedding methods. A similar performance trend to that in Tables~\ref{pesq} and~\ref{estoi} is observed.}

\begin{table}[tbp]
\vspace{-1.5em}
\centering
    \scriptsize
    \def\arraystretch{0.93}
    \setlength{\tabcolsep}{1.55pt}
    \setlength{\abovetopsep}{0pt}
    \setlength\belowbottomsep{0pt} 
    \setlength\aboverulesep{0pt} 
    \setlength\belowrulesep{0pt}
\caption{Average CSIG, CBAK, and COVL scores. The best scores are in boldface.}
\vspace*{0.05in}
\scalebox{1}{
\begin{tabular}{l|c|c|ccc|ccc}
\toprule
\multirow{2}{*}{\makecell[c]{\textbf{Methods}}} 
& \multirow{2}{*}{\makecell[c]{\textbf{Types}}}
& \multirow{2}{*}{\makecell[c]{\textbf{Causality}}}
& \multicolumn{3}{c|}{\textbf{MS}} 
& \multicolumn{3}{c}{\textbf{PSM}}  \\
\cmidrule{4-9}
&  &  & CSIG & CBAK & COVL & CSIG & CBAK & COVL \\
\midrule
Noisy & -- & -- & 2.26  & 1.80  & 1.67  & -- & -- & --  \\
\midrule

No-Pos & -- & \multirow{5}{*}{\textbf{Yes}} & 

\textcolor{black}{3.17}  & \textcolor{black}{2.38}  & \textcolor{black}{2.46}  &

\textcolor{black}{3.18} & \textcolor{black}{2.54} & \textcolor{black}{2.48}  \\

\makecell[c]{Sinusoidal-APE~\cite{transformer}} 
& \multirow{2}{*}{APE} & 

& \textcolor{black}{3.16} & \textcolor{black}{2.38} & \textcolor{black}{2.46}

& \textcolor{black}{3.20} & \textcolor{black}{2.54} & \textcolor{black}{2.49} \\
Learned-APE~\cite{bert} 
&  &
& \textcolor{black}{3.16} & \textcolor{black}{\textbf{2.40}} & \textcolor{black}{2.46}

& \textcolor{black}{3.20} & \textcolor{black}{2.53} & \textcolor{black}{2.50} \\
T5-RPE~\cite{T5} 
& \multirow{2}{*}{RPE} &

& \textcolor{black}{\textbf{3.20}} & \textcolor{black}{\textbf{2.40}} & \textcolor{black}{\textbf{2.49}}

& \textcolor{black}{\textbf{3.21}} & \textcolor{black}{2.54} & \textcolor{black}{\textbf{2.51}} \\

KERPLE~\cite{chi2022kerple} 
& &

& \textcolor{black}{3.16} & \textcolor{black}{\textbf{2.40}} & \textcolor{black}{2.46}

& \textcolor{black}{3.20} & \textcolor{black}{\textbf{2.55}} & \textcolor{black}{\textbf{2.51}} \\

\midrule
\midrule

No-Pos & -- & \multirow{5}{*}{\textbf{No}} &

\textcolor{black}{3.16}  & \textcolor{black}{2.35}  & \textcolor{black}{2.44}  &

\textcolor{black}{3.16} & \textcolor{black}{2.54} & \textcolor{black}{2.47}  \\

\makecell[c]{Sinusoidal-APE~\cite{transformer}} 
& \multirow{2}{*}{APE} & 

& \textcolor{black}{3.31} & \textcolor{black}{2.50} & \textcolor{black}{2.63} 

& \textcolor{black}{3.35} & \textcolor{black}{2.65} & \textcolor{black}{2.65} \\
Learned-APE~\cite{bert} 
&  &
& \textcolor{black}{3.26} & \textcolor{black}{2.48} & \textcolor{black}{2.58}

& \textcolor{black}{3.32} & \textcolor{black}{2.65} & \textcolor{black}{2.61} \\
T5-RPE~\cite{T5}  
& \multirow{2}{*}{RPE} &

& \textcolor{black}{\textbf{{3.35}}} & \textcolor{black}{\textbf{{2.54}}} & \textcolor{black}{\textbf{{2.68}}}

& \textcolor{black}{\textbf{3.40}} & \textcolor{black}{\textbf{2.70}} & \textcolor{black}{\textbf{2.70}} \\

KERPLE~\cite{chi2022kerple}
& &

& \textcolor{black}{3.32} & \textcolor{black}{2.50} & \textcolor{black}{2.63}

& \textcolor{black}{3.35} & \textcolor{black}{2.66} & \textcolor{black}{2.65} \\

       
\toprule

\end{tabular}}
\label{composite}
\vspace{-2.3em}
\end{table}

\vspace{-0.5em}
\section{CONCLUSION}\label{sec:6}
\vspace{-0.3em}

In this paper, we present a systematic empirical study that investigates the impact of position embedding on speech enhancement using Transformers. Our comprehensive experiments encompass two commonly used training targets, i.e., MS (spectral mapping) and PSM (spectral masking). The results clearly indicate that position embeddings often limited benefits for Transformer-based speech enhancement in a causal configuration. Conversely, they significantly enhance performance in a noncausal configuration. Furthermore, our comparative analysis underscores the superiority of relative position encoding (RPE) over absolute position encoding (APE).

\newpage
\newpage

\vfill\pagebreak

\bibliographystyle{IEEEtran}
\begin{footnotesize}
\bibliography{refs}
\end{footnotesize}








\end{document}